\documentclass[prl,twocolumn,showpacs,preprintnumbers,amsmath,amssymb]{revtex4}

\usepackage{graphicx}
\usepackage{dcolumn}
\usepackage{bm}
\usepackage{psfrag}
\usepackage{epsfig}
\usepackage{amsmath}
\usepackage{amssymb}
\usepackage{color}

\newcommand{\nn}{\nonumber}
\newcommand{\Bra}{\langle}
\newcommand{\Ket}{\rangle}

\newcommand{\ignore}[1]{}

\begin{document}

\title{Polariton crystallization in driven arrays of lossy nonlinear resonators}

\author{Michael J. Hartmann}
\email{michael.hartmann@ph.tum.de}
\affiliation{Technische Universit{\"a}t M{\"u}nchen, Physik Department, James-Franck-Str., 85748 Garching, Germany}

\date{\today}

\begin{abstract}
We investigate the steady states of a lossy array of nonlinear optical resonators that are driven by lasers and interact
via mutual photon tunneling. For weak nonlinearities, we find two-mode squeezing of polaritons in modes whose quasi-momenta
match the relative phases of the laser drives. For strong nonlinearities the spatial polariton density-density correlations
indicate that the polaritons crystallize and are predominantly found at a specific distance from each other despite being injected by a coherent light source and damped by the environment.
\end{abstract}

\maketitle

%
%
\paragraph{Introduction --}
Interacting quantum many-body systems~\cite{BDZ08} give rise to a number of fascinating phenomena
such as quantum phase transitions, quantum magnetism or charge fractionalization.
Of particular interest is the strongly correlated regime, where collective phenomena are most pronounced.
In most cases however, strongly correlated quantum many-body systems are studied in scenarios of thermodynamic equilibrium,
that permit a description with statistical techniques, and substantial understanding of, e.g. equilibrium quantum phase transitions~\cite{Sachdevbook}, has been achieved.
On the other hand, a lot less is known about non-equilibrium regimes where the balance between loading and loss mechanisms leads to the emergence of stationary states. Here we investigate collective phenomena in non-equilibrium steady states of lossy
arrays of coupled nonlinear optical resonators that are coherently driven by lasers.

Strongly interacting polaritons~\cite{HBP06,HBP08} and photons in coupled arrays of micro-cavities~\cite{ASB06} and optical fibers~\cite{CGM+08,KH09}
have recently been shown to be suitable candidates for realizing a strongly correlated many-body regime with current technology~\cite{ADW+06,BHB+09}.
So far, possibilities to observe equilibrium phenomena, such as a Mott insulator~\cite{HBP06,ASB06} or a Tonks-Girardeau gas~\cite{CGM+08,KH09}, have mostly been addressed. These regimes have however been realized previously, e.g. with ultra cold atoms~\cite{BDZ08}. In contrast, we here predict a phenomenon for polaritons, for which no analogue in other implementations is known so far.

Experiments to generate quantum states with photons, either in cavity QED~\cite{ADW+06} or with optical fibers~\cite{BHB+09}, typically work in non-equilibrium situations and it is therefore much more natural and feasible to consider driven dissipative scenarios.
First steps in this direction have been undertaken with studies of an optical Josephson effect~\cite{GTI+09},
the dynamical evolution for nonlinearities initially prepared in a non-equilibrium state~\cite{TGF+09}, an analysis of the spectroscopical properties of driven dissipative nonlinearities~\cite{CGT+09} and entanglement studies~\cite{AMB07}.

In this work, we consider arrays of cavities that are driven by lasers of constant intensity and dissipate photons into their environment. Photons can tunnel between neighboring cavities and interact with suitable emitters in each cavity in such a way that they
form polaritons and experience an optical nonlinearity.
In this scenario, the interplay of laser drive and photon loss leads to the emergence of steady states, for which we derive the particle statistics and characteristic correlations.

We find two main results. In the regime, where the Rabi frequencies of the driving lasers are much stronger than the nonlinearities, only one Bloch mode with quasi-momentum $k$ is driven by the lasers, and we find two-mode squeezing for modes with quasi-momenta $p$ and $\bar{p}$, such that $p + \bar{p} = k$.
Since this sqeezing emerges for weak nonlinearities, an experimental observation would not require a strong coupling regime for the employed cavities.
In the complementary regime, where the nonlinearities are much stronger than the laser drives, we find spatial anti-correlations of the polariton densities indicating that polaritons crystallize and are predominantly found within a specific distance from one another. As it requires strong nonlinearities, the crystallization can only be generated in cavities that operate in the strong coupling regime~\cite{ADW+06}. We stress that this polariton crystallization appears despite the fact that coherent lasers continuously drive the cavities and damping permanently dissipates photons. The emergence of crystallization in the dissipative scenario with coherent drive is the most significant result of this work and has no analogue in other realizations.

\paragraph{Model --}

Since bare photons do not interact, photon-photon interactions or optical nonlinearities only emerge when light interacts with optical emitters. Depending on the strength of the photon-emitter coupling, the elementary excitations of the system are either photons, for weak coupling, or polaritons, superpositions of photons and emitter excitations, for strong coupling. In the following we will use the term ``polaritons'' for both regimes.
For both regimes, their Hamiltonian can be taken to read,
%
\begin{align} \label{eq:ham}
H & = \Delta \sum_{j=1}^{N} a_{j}^{\dagger} a_{j} - J \sum_{j=1}^{N} \left( a_{j}^{\dagger} a_{j+1} + a_{j} a_{j+1}^{\dagger} \right) \\
& + \frac{U}{2} \sum_{j=1}^{N} a_{j}^{\dagger} a_{j}^{\dagger} a_{j} a_{j} + \sum_{j=1}^{N} \left( \frac{\Omega_{j}}{2} a_{j}^{\dagger} + \frac{\Omega_{j}^{\star}}{2}a_{j} \right),\nn
\end{align}
in a frame that rotates at the frequency $\omega_{L}$ of the driving lasers (we set $\hbar = 1$). We assume periodic boundary conditions, the index $j$ labels the resonators and $a_{j}^{\dagger}$ ($a_{j}$)  creates (annihilates) a polariton in resonator $j$.
Polaritons interact with strength $U$ in each resonator and tunnel between neighboring resonators at rate $J$. $\Delta = \omega_{pol}-\omega_{L}$ is the detuning between polariton and laser frequency and $\Omega_{j}$ are the Rabi frequencies of the driving lasers. 
We assume that all lasers have the same amplitude, but may have different phases,
$\Omega_{j} = \Omega e^{- i \phi_{j}}$. Only relative phases of the lasers matter and we can choose $\Omega > 0$. For now, we choose $\phi_{j} = \frac{\pi}{2} j$ ($j = 1, 2, \dots, N$) and $N$ to be a multiple of $4$ for reasons that will become clear in the sequel. Other values of $\phi_{j}$ will be considered below. 

The Hamiltonian (\ref{eq:ham}) can be implemented in several ways~\cite{HBP08}. One suitable approach~\cite{HBP06} makes use of dark state polaritons in 4-level atoms,
where a dispersively operated two polariton process gives rise to the interaction term $\frac{U}{2} a_{j}^{\dagger} a_{j} ( a_{j}^{\dagger} a_{j} - 1)$ in each resonator.
The dynamics of the system, including polariton losses from the cavities at a rate $\gamma$ is given by the master equation
\begin{equation}\label{eq:master}
\dot{\rho} = - i [H,\rho] + \frac{\gamma}{2} \sum_{j=1}^{N} \left( 2 a_{j} \rho a_{j}^{\dagger} - a_{j}^{\dagger} a_{j} \rho - \rho a_{j}^{\dagger} a_{j} \right)\, .
\end{equation}
The Hamiltonian $H$ of eq. (\ref{eq:ham}) can also be written in terms of Bloch modes,
$B_{k} = \frac{1}{\sqrt{N}} \sum_{j=1}^{N} e^{i k j} a_{j}$,
where $k = \frac{2 \pi l}{N}$ and $l = -\frac{N}{2} + 1,-\frac{N}{2} + 2,\dots,\frac{N}{2}$, to read
$
H = \sum_{k} \omega_{k} B_{k}^{\dagger} B_{k} + \frac{\sqrt{N} \Omega}{2} (B_{\frac{\pi}{2}} + B_{\frac{\pi}{2}}^{\dagger} )
+ \frac{U}{2 N} \sum_{k_{1},k_{2},k_{3},k_{4}} \delta_{k_{1} + k_{2} + 2 \pi z,k_{3} + k_{4}}
B_{k_{1}}^{\dagger} B_{k_{2}}^{\dagger} B_{k_{3}} B_{k_{4}}
$
with an arbitrary integer $z$ and $\omega_{k} = \Delta - 2 J \cos k$. The damping terms transform to
$\sum_{k} ( 2 B_{k} \rho B_{k}^{\dagger} - B_{k}^{\dagger} B_{k} \rho - \rho B_{k}^{\dagger} B_{k} )$.

For our specific choice of $N$ and $\phi_{j} = \frac{\pi}{2} j$, lasers that drive each cavity resonantly ($\Delta = 0$), constructively interfere in driving the mode $B_{\frac{\pi}{2}}$ of the same frequency $\omega_{\frac{\pi}{2}} = \Delta = 0$. Lasers that are in phase,
$\phi_{j} = \phi_{0}$, would destructively interfere for this mode, $B_{\frac{\pi}{2}}$, thus motivating our choice of $N$ and $\Omega_{j}$. We note that the lasers generate a polariton flow in the cavity array, that can roughly be estimated as $\sim J \sin \phi$, where $\phi = i \ln (\Omega_{j+1}/\Omega_{j})$ is phase difference between the driving lasers of adjacent cavities, and becomes maximal for $\phi =\pi/2$.
We now analyze the steady states of eq. (\ref{eq:master}), for which $\dot{\rho} = 0$.

\paragraph{Strong driving regimes --}

\begin{figure} 
\includegraphics[width=4.2cm]{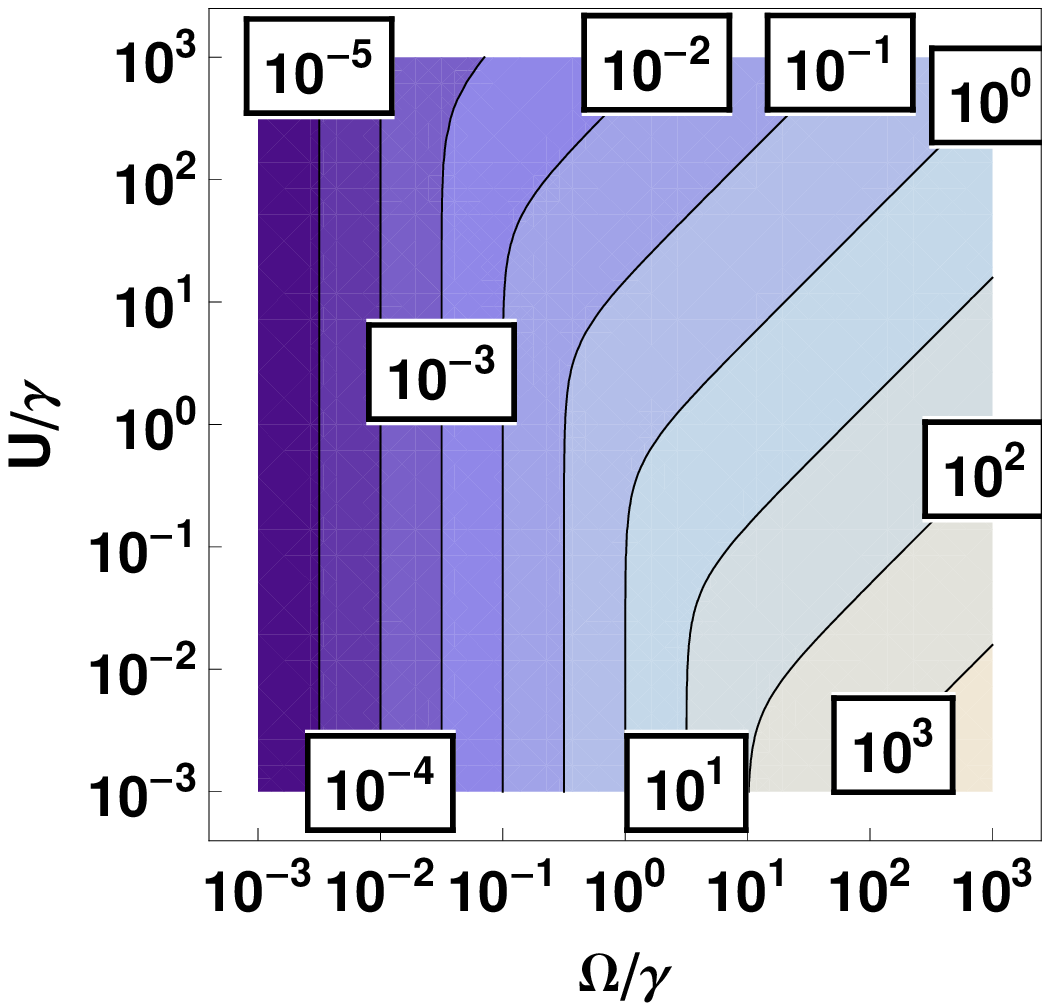}
\hspace{0.02cm}
\includegraphics[width=4.2cm]{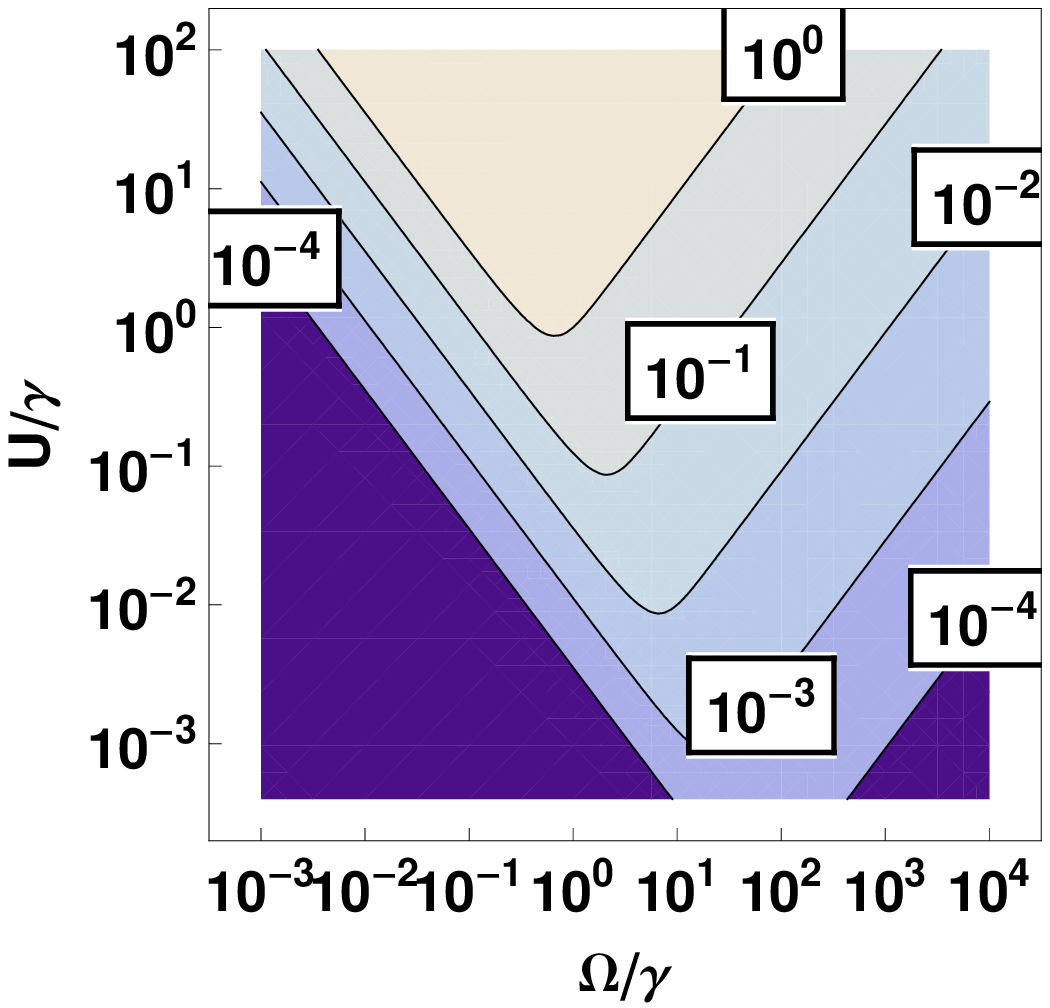}
\caption{\label{strongdriveplot} The steady state in the strong driving regime. Left: $n$ as given by eq (\ref{eq:nvalue}) as a function of $\Omega/\gamma$ and $U/\gamma$. Right: $m/n$ as given by eqs. (\ref{eq:moments}) and (\ref{eq:nvalue}) as a function of $\Omega/\gamma$ and $U/\gamma$.}
\end{figure}

The $k=\pi/2$ mode is, in contrast to all other modes driven by the lasers and polaritons from this mode can only scatter into other modes via the nonlinearities $U$. For regimes, where $\Omega \gg U$, one thus expects that the state of the polariton field in the cavity array can be well approximated by a coherent state in the mode $k=\pi/2$ plus small perturbations.
We therefore split the mode operators, $B_{k} = \beta_{k} + b_{k}$, into coherent parts, represented by a complex number $\beta_{k}$ and quantum fluctuations $b_{k}$, where $\beta_{k} = \sqrt{N} \beta \delta_{k,\frac{\pi}{2}}$.
Neglecting all quantum fluctuations, $b_{k}$, the background field $\beta$ obeys the equation of motion,
$\dot{\beta} = -i \frac{\Omega}{2} - i U |\beta|^{2} \beta - \frac{\gamma}{2} \beta$,
and for the steady state, the density of photons in the background field, $n = |\beta|^{2}$, is determined by the equation
$
4 U^{2} n^{3} + \gamma^{2} n = \Omega^{2}
$,
which has
\begin{equation} \label{eq:nvalue}
n = (3^{1/3} X^{2/3}-3^{2/3} \gamma ^2)/(6 U X^{1/3})
\end{equation}
with $X = 9 U \Omega ^2+\sqrt{3} \sqrt{\gamma ^6+27 U^2 \Omega ^4}$ as the only real and positive solution. Furthermore, a stability analysis~\cite{DW80} shows, that this solution is always stable, which guarantees the existence of a unique steady state. The left plot in figure~\ref{strongdriveplot} shows $n$ as a function of $U/\gamma$ and $\Omega/\gamma$. $n$ is maximal for $U=0$ and $\Omega \gg \gamma$.
Expanding the Hamiltonian to second order in $b_{k}$ and $b_{k}^{\dagger}$ we obtain,
\begin{equation} \label{eq:hamfourier-pert}
H = \sum_{k} \left[ \left(\omega_{k} + 2 U n \right) b_{k}^{\dagger} b_{k}
+ \left( \frac{U \beta^{2}}{2} b_{k}^{\dagger} b_{\bar{k}}^{\dagger} + \text{h.c.} \right) \right] \, ,
\end{equation}
where $\bar{k} = \frac{k}{|k|} \pi - k$ and terms that are linear in $b_{\frac{\pi}{2}}$ have been neglected as they cancel in the corresponding master equation by virtue of eq. (\ref{eq:nvalue}).
The Hamiltonian (\ref{eq:hamfourier-pert}) is known to lead to two-mode squeezing for the pairs of modes ($k$, $\bar{k}$)~\cite{WMbook}. The corresponding master equation (\ref{eq:master}) with $H$ as in eq. (\ref{eq:hamfourier-pert}) is quadratic in the operators $b_{k}^{\dagger}$ and $b_{k}$. Its steady state is therefore a Gaussian state,
that is completely determined by the first and second order moments of $b_{k}^{\dagger}$ and $b_{k}$, which are zero except for,
$\Bra b_{k}^{\dagger} b_{k} \Ket = m$ and $\Bra b_{k} b_{\bar{k}} \Ket = g$, where
\begin{equation} \label{eq:moments}
m = \frac{2 U^{2} n^{2}}{12 U^{2} n^{2} + \gamma^{2}}, \quad
g = - \frac{4 U^{2} n + i U \gamma}{12 U^{2} n^{2} + \gamma^{2}} \beta^{2}.
\end{equation}
We can now check the validity of our approximation by verifying that $\sum_{k} \Bra b_{k}^{\dagger} b_{k} \Ket \ll N \, n \, \Leftrightarrow m \ll n$.
The resulting phase diagram is shown in the right plot of figure~\ref{strongdriveplot}, where we plot $m/n$ as a function of $U/\gamma$ and $\Omega/\gamma$.
In the regime with $\Omega > \gamma$, there is on average more than one polariton in each cavity, $n > 1$, and for increasing nonlinearities $U$, the state differs significantly from a coherent state in mode $k = \pi/2$. For $\Omega < \gamma$, on the other hand, the polariton density is small, $n < 1$. Since the nonlinearities only affect states with more than one polariton, they become ineffective in this regime and the state remains coherent for higher values of $U$.
  
To obtain a more detailed picture of the steady state we study its particle statistics, which can be calculated via its characteristic function~\cite{WMbook}.
For the first order coherence between modes $k$ and $p$ we obtain $\Bra B_{k}^{\dagger} B_{p} \Ket = \delta_{k,\frac{\pi}{2}} \delta_{p,\frac{\pi}{2}} N n + \delta_{k,p} m$, whereas the density-density correlations between modes $k$ and $p$, $g_{\text{m}}^{(2)}(k,p) = \Bra B_{k}^{\dagger} B_{p}^{\dagger} B_{p} B_{k} \Ket/(\Bra B_{k}^{\dagger} B_{k} \Ket \Bra B_{p}^{\dagger} B_{p} \Ket)$, read,
$g_{\text{m}}^{(2)}(k,p) \approx 1 +  \delta_{k\not=\frac{\pi}{2}} \left(\delta_{k,p} + \delta_{k+p,\pm \pi} \frac{|g|^{2}}{m^{2}}\right) - \delta_{k,\frac{\pi}{2}} \delta_{p,\frac{\pi}{2}} \frac{2 m}{N n}$
to leading order in $1/(N n)$. Here we have taken into account that $|g|^{2}, m^{2} \ll n^{2}$. Due to the weak nonlinearity $U$, the driven mode $k = \frac{\pi}{2}$ shows slight anti-bunching $g_{\text{m}}^{(2)}(\frac{\pi}{2},\frac{\pi}{2}) \approx 1 - \frac{2 m}{N n}$~\cite{WMbook}.
The nonlinearity $U$ always scatters two polaritons from mode $k = \frac{\pi}{2}$ synchronously into modes $k$ and $\bar{k}$, c.f. eq. (\ref{eq:hamfourier-pert}). This leads to two-mode squeezing where polariton pairs as given by $g$ in eq. (\ref{eq:moments}) are created.
The term $\delta_{k+p,\pm \pi} \frac{|g|^{2}}{m^{2}}$ describes correlations that originate from these pairs.
Since $\frac{|g|^{2}}{m^{2}} = 4 + \gamma^{2}/(4 n^{2} U^{2})$, these are most pronounced for strong damping, $U/\gamma \ll1$ and $\Omega/\gamma \ll 1$, where however the intensity becomes increasingly weak, $n \ll 1$, see fig. \ref{strongdriveplot}. 
Importantly the observation of these pairing correlations does not require a strong coupling regime for the employed cavities.
Nonetheless, the effect is not classical and disappears for $U \rightarrow 0$, where $g \rightarrow 0$, c.f. eq. (\ref{eq:moments}).

For the correlations between different resonators, $j$ and $l$, $g_{\text{r}}^{(1)}(j,l) = \Bra a_{j}^{\dagger} a_{l} \Ket / \sqrt{\Bra a_{j}^{\dagger} a_{j} \Ket \Bra a_{l}^{\dagger} a_{l} \Ket}$ and $g_{\text{r}}^{(2)}(j,l) = \Bra a_{j}^{\dagger} a_{l}^{\dagger} a_{l} a_{j} \Ket / (\Bra a_{j}^{\dagger} a_{j} \Ket \Bra a_{l}^{\dagger} a_{l} \Ket)$, we find for $|g|^{2}, m^{2} \ll n^{2}$,
$g_{\text{r}}^{(1)}(j,l) \approx e^{i \frac{\pi}{2} (j-l)}$, $g_{\text{r}}^{(2)}(j,l) \approx 1 - 2 \frac{m}{n} \delta_{j,l}$,
and hence, as expected for a nonlinearity, a spatial anti-bunching of $g_{\text{r}}^{(2)}(j,j) \approx 1 - 2 \frac{m}{n}$~\cite{WMbook}.

It is important to note, that all above results are independent of the tunneling rate $J$, a feature that only appears for our specific choice, $\phi_{j} = \frac{\pi}{2} j$, for the phases of the lasers. For laser phases where $|\Delta - J \cos (\phi_{j}-\phi_{j-1})| > \frac{\sqrt{3}}{2} \gamma$, bistabilities of the steady state appear. A detailed discussion of these cases will be presented elsewhere.

\paragraph{Weak driving regimes --}
We now analyze the regime, where the Rabi frequencies of the driving lasers $\Omega_{j}$ are weaker than the nonlinearities $U$.
For these parameters we have to resort to numerical calculations. We represent the density matrix of the polaritons as a Matrix Product Operator and employ a TEBD algorithm~\cite{ZV04} that integrates the master equation (\ref{eq:master}) in time until a steady state is reached. We use a 2nd order Trotter decomposition for the Lindblad super-operator with time steps $\delta t = U/100$ (or $\delta t = U/50$) and keep relative errors due to matrix truncation below $10^{-8}$ (or $10^{-6}$) at each time step allowing for matrix dimension up to $300 \times 300$. We consider an array of 16 cavities and, since $\Omega \ll U$, truncate the Hilbert space to allow for up to 2 polaritons in each cavity. 

In a first example we choose $\Delta = 0$, $U/\gamma = 10$, $\Omega/\gamma = 2$ and $J/\gamma = 1$.
\begin{figure}
\centering
\includegraphics[width=9cm]{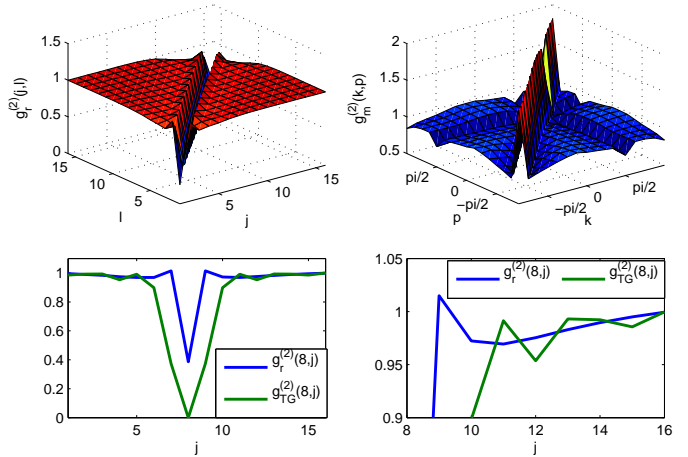}
\caption{\label{fig:G01Om02} Density correlations of the steady state for 16 cavities in the weak driving regime with $\Delta = 0$, $U/\gamma = 10$, $\Omega/\gamma = 2$ and $J/\gamma = 2$. Top left: $g_{\text{r}}^{(2)}(j,l)$. Top right: $g_{\text{m}}^{(2)}(k,p)$. Bottom left: $g_{\text{r}}^{(2)}(8,l)$  and $g_{\text{TG}}^{(2)}(8,l)$. Bottom right: $g_{\text{r}}^{(2)}(8,l)$  and $g_{\text{TG}}^{(2)}(8,l)$, zoomed in on $8 \le j \le 16$.}
\end{figure}
Figure \ref{fig:G01Om02} shows density correlations, $g_{\text{r}}^{(2)}(j,l)$ (top left) and $g_{\text{m}}^{(2)}(k,p)$ (top right), for the steady state.
As a consequence of the strong nonlinearity there is a pronounced anti-bunching, $g^{(2)}(j,j) \ll 1$~\cite{WMbook}. In marked contrast to the strong driving regimes, polaritons in the same mode are strongly paired, $g_{\text{m}}^{(2)}(k,k) > 1$, but polaritons in different modes are
strongly anti-correlated or anti-paired, $g_{\text{m}}^{(2)}(k,p) < 1$ for $k \not= p$. That is, if a polariton is found in mode $k$, the probability to find a second polariton in a mode $p \not= k$ is lower than for independent particles.
$g_{\text{m}}^{(2)}(k,p)$ is smallest for $k \not= p$, where one quasi-momentum equals $\pi / 2$.

Most interestingly, $g^{(2)}(j,j+1)$ is larger than unity whereas $g^{(2)}(j,j+2)$ and $g^{(2)}(j,j+3)$ etc. are significantly below unity,
see bottom row of figure \ref{fig:G01Om02}.
Polariton densities in neighboring cavities are thus correlated whereas they are anti-correlated for larger separations. That is, if a polariton is found in one cavity, the probability to find a second polariton is for separation 1 higher and for larger separations lower than for independent particles. This behavior indicates that the polaritons are crystallized and predominantly occur at distances of one cavity-cavity separation from each other. More specifically, the polaritons form dimers that are extended across two neighboring resonators and move along the array due to the flow created by the relative phases of the lasers. 

Since our system is one-dimensional, has significantly less than one polariton per cavity and polaritons in the same cavity strongly interact, one might be tempted to compare it to a Tonks-Girardeau gas~\cite{BDZ08}. For the latter, $g^{(2)}$ is the same as for free fermions and shows oscillating anti-correlations, which are known as Friedel oscillations~\cite{F58} and appear as a consequence of the Pauli exclusion principle.
The density anti-correlations we find are quantitatively different from Friedel oscillations of free Fermi or Tonks-Girardeau gases,
$g_{\text{TG}}^{(2)}(j,l) = 1 - (\frac{\sin(\pi \tilde{n} (j-l))}{\pi \tilde{n} (j-l)})^{2}$,
where $\tilde{n}$ is the number of polaritons per cavity. This is shown in the bottom row of fig. \ref{fig:G01Om02}, where we plot
$g_{\text{r}}^{(2)}(8,j)$ and $g_{\text{TG}}^{(2)}(8,j)$ with $\tilde{n} = \tilde{n}(8)$. In contrast to $g_{\text{r}}^{(2)}$, $g_{\text{TG}}^{(2)}$ only shows anti-correlations, $g_{\text{TG}}^{(2)} \le 1$. Whereas the amplitude of the anti-correlations is comparable, they do not oscillate.

Next we consider the dependence of the densities, $\tilde{n}$, and density correlations, $g^{(2)}(j,l)$, on the polariton tunneling $J$ in more detail.
\begin{figure}
\centering
\includegraphics[width=8cm]{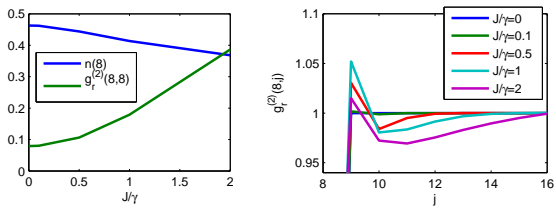}
\caption{\label{fig:g2undnvonJ} $g_{\text{r}}^{(2)}(8,j)$ and $\tilde{n}(8)$ as functions of $J/\gamma$. Left: $g_{\text{r}}^{(2)}(8,8)$ and $\tilde{n}(8)$ as functions of $J/\gamma$. Right $g_{\text{r}}^{(2)}(8,j)$ for $8 \le j \le 16$ and $J/\gamma = 0, 0.1, 0.5, 1$ and $2$. $\Delta = 0$, $U/\gamma = 10$ and $\Omega/\gamma = 2$.}
\end{figure}
We have computed $g^{(2)}(8,j)$ and $\tilde{n}(8)$ for $J/\gamma = 0, 0.1, 0.5, 1$ and $2$, where the other parameters are $\Delta = 0$, $U/\gamma = 10$ and $\Omega/\gamma = 2$. The left plot of figure \ref{fig:g2undnvonJ} shows $g_{\text{r}}^{(2)}(8,8)$ and $\tilde{n}(8)$ as functions of $J/\gamma$, whereas the right plot shows $g_{\text{r}}^{(2)}(8,j)$ for $8 \le j \le 16$ and $J/\gamma = 0, 0.1, 0.5, 1$ and $2$. The crystallization signatures appear for nonzero tunneling $J$ only and density anti-correlations become increasingly pronounced and long ranged as $J$ is increased.  

To confirm the experimental robustness of our findings we have computed $g_{\text{r}}^{(2)}(8,j)$ for smaller nonlinearities,  $U$, and various phase differences, $\phi = i \ln (\Omega_{j+1}/\Omega_{j})$.
\begin{figure}
\centering
\includegraphics[width=8cm]{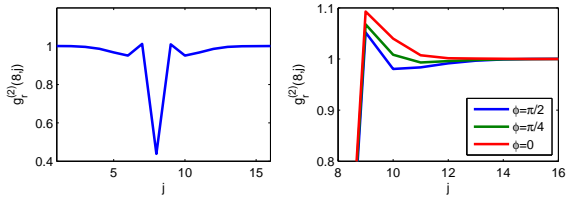}
\caption{\label{fig:dampandphases} Left: $g_{\text{r}}^{(2)}(8,j)$ for $\Delta = 0$, $J/\gamma = 1$, $\Omega/\gamma = 2$ and $U/\gamma = 5$. Right: $g_{\text{r}}^{(2)}(8,j)$ for $\Delta = 0$, $U/\gamma = 10$, $J/\gamma = 1$ and $\Omega/\gamma = 2$ for $\phi = 0, \pi/4$ and $\pi/2$.}
\end{figure}
Figure \ref{fig:dampandphases} shows $g_{\text{r}}^{(2)}(8,j)$ for $\Delta = 0$, $J/\gamma = 1$ and $\Omega/\gamma = 2$.
In the left plot we choose $\phi = \pi/2$ and set $U/\gamma = 5$. In the right plot we choose $U/\gamma = 10$ and  consider $\phi = 0, \pi/4$ and $\pi/2$. Anti-correlations only appear for $\phi \not= 0$.

Whereas the strong anti-bunching, $g_{\text{r}}^{(2)}(j,j) \ll 1$, is expected for a strong nonlinearity~\cite{WMbook}, the anti-correlations, $g_{\text{r}}^{(2)}(j,l) < 1$ for $|j-l| \ge 2$, are more surprising. They emerge due to and interplay between the nonlinearities and the polariton flow generated by the relative phase differences of the lasers, which in turn is scattered at the nonlinearities. 

In all our examples we find $\tilde{n}(j) < 0.5$ and $g^{(2)}(j,j) \ll 1$, which confirms that truncating the local Hilbert space to states of at most 2 polaritons is indeed a good approximation. The validity of our calculations is also substantiated by their excellent agreement with the exact analytical solution for the $J=0$ limit~\cite{DW80}.

\paragraph{Experimental realization and measurements --}
The crystallization of polaritons we predicted here can be observed with resonators of high single emitter cooperativity, such as microtoroids, circuit cavities, photonic band gap cavities, micropillar Bragg stacks or Fabry-P\'erot microcavities on a silicon chip~\cite{ADW+06}. A straight forward method to measure the correlations we derived is to detect the light emitted from the structure. Detection of near-field photons with detectors of sufficiently fast response time gives access to correlations between cavities, $g_{\text{r}}^{(1)}(j,l)$ and $g_{\text{r}}^{(2)}(j,l)$ whereas the far-field carries information on correlations between the modes, $g_{\text{m}}^{(1)}(k,p)$ and $g_{\text{m}}^{(2)}(k,p)$. 
Furthermore, the polaritons are superpositions of photons and emitter excitations and their statistics and correlations can be inferred from measurements on the emitters. In some implementations, the polaritons can even be perfectly transferred onto the emitters prior to the measurement~\cite{HBP06}. Even though the variations of $g_{\text{r}}^{(2)}(j \ne l)$ are only in the range $0.95 \le g_{\text{r}}^{(2)}(j \ne l) \le 1.05$, they can reliably be measured since $g_{\text{r}}^{(2)}(j \ne l)$ is a ratio of density correlations which are both affected by detector inefficiencies in the same way, leaving their ratio unaffected.

\begin{acknowledgments}
The author thanks M. Kiffner and A. Recati for discussions. 
This work is part of the Emmy Noether project 
HA 5593/1-1 funded by Deutsche Forschungsgemeinschaft (DFG).
\end{acknowledgments}


\begin{thebibliography}{99}
%
\bibitem{BDZ08}
I.~Bloch, J.~Dalibard and W.~Zwerger,
Rev. Mod. Phys. {\bf 80}, 885 (2008).
%
\bibitem{Sachdevbook}
S.~Sachdev, {\it Quantum Phase Transitions}, Cambridge University Press 1999.
%
\bibitem{HBP06}
M.J.~Hartmann, F.G.S.L.~Brand\~ao and M.B.~Plenio, Nat. Phys. {\bf 2}, 849 (2006);
M.J. Hartmann and M.B. Plenio, Phys. Rev. Lett. {\bf 99}, 103601 (2007)
%
\bibitem{HBP08}
M.J.~Hartmann, F.G.S.L.~Brand\~ao and M.B.~Plenio, Laser \& Photon. Rev. {\bf 2}, 527 (2008).
%
\bibitem{ASB06}
D.G.~Angelakis, M.F. Santos and S. Bose, Phys. Rev. A {\bf 76}, 031805(R) (2007);
A.D. Greentree et~al., Nat. Phys. {\bf 2}, 856~(2006)
%
\bibitem{KH09}
M.~Kiffner and M.J.~Hartmann, arXiv:0908.2055
%
\bibitem{CGM+08}
D.~E.~Chang et~al., Nat. Phys. {\bf 4}, 884 (2008).
%
\bibitem{ADW+06}
T.~Aoki et~al., Nature {\bf 443}, 671 (2006);
K.~Hennessy et~al., Nature {\bf 445}, 896 (2007);
M.~Trupke et~al., Phys. Rev. Lett. {\bf 99}, 063601 (2007);
A.~Wallraff et~al., Nature {\bf 431}, 162 (2004);
J.~P.~Reithmaier et~al., Nature {\bf 432}, 197 (2004).
%
\bibitem{BHB+09}
M.~Bajcsy et~al., Phys. Rev. Lett. {\bf 102}, 203902 (2009).
%
\bibitem{GTI+09}
D.~Gerace et~al., Nat. Phys. {\bf 5}, 281 (2009);
%
\bibitem{CGT+09}
I.~Carusotto et~al., Phys. Rev. Lett. {\bf 103}, 033601 (2009).
%
\bibitem{TGF+09}
A.~Tomadin et~al.,
arXiv:0904.4437
%
\bibitem{AMB07}
M.~B.~Plenio and S.~F.~Huelga
Phys. Rev. Lett. {\bf 88}, 197901 (2002);
D.~G.~Angelakis, S.~Mancini and S.~Bose,
Europhys. Lett. {\bf 85}, 20007 (2009);
D.~G.~Angelakis, L.~Dai and L.-C.~Kwek,
arXiv:0906.2168.
%
\bibitem{DW80}
P.D. Drummond and D.F. Walls,
J. Phys. A: Math. Gen., {\bf 13}, 725 (1980)
%
\bibitem{F58}
J.~Friedel,
Nuovo Cimento {\bf 7}, 287 (1958).
%
%
\bibitem{ZV04}
M.~Zwolak and G.~Vidal,
Phys. Rev. Lett. {\bf 93}, 207205 (2004);
U.~Schollw\"ock, Rev. Mod. Phys. {\bf 77}, 259 (2005).
%
\bibitem{WMbook}
D.F. Walls and G. Milburn, {\it Quantum Optics}, Springer 2007.
%
\end{thebibliography}
\end{document}